\documentclass[usenatbib]{basi}
\usepackage[T1]{fontenc}
\usepackage[british]{babel}
\usepackage[varg]{txfonts}
\usepackage{rotating}
\usepackage{dcolumn}
\begin{document}
\title[Galaxies in 21-cm absorption]{Galaxies in H~{\sc i} 21-cm absorption at $z<$3.5}
\author[N. Gupta et~al.]%
       {N. Gupta$^1$\thanks{email: \texttt{ngupta@iucaa.ernet.in}},
       R. Srianand$^{1}$, P. Petitjean$^2$, P. Noterdaeme$^2$ and E. Momjian$^3$\\
       $^1$IUCAA, Post Bag 4, Ganeshkhind, Pune 411007, India\\
       $^2$UPMC-CNRS, UMR 7095, Institut d'Astrophysique de Paris, 75014 Paris, France\\
       $^3$National Radio Astronomy Observatory, 1003 Lopezville Road, Socorro, NM 87801, USA
       }

\pubyear{2014}
\volume{00}
\pagerange{\pageref{firstpage}--\pageref{lastpage}}

\date{Received --- ; accepted ---}

\maketitle
\label{firstpage}

\begin{abstract}
We present recent results from our searches of 21-cm absorption using GBT, GMRT, VLBA and WSRT 
to trace the evolution of cold gas in galaxies.  
Using $\sim$130 sight lines with 21-cm absorption measurements, we find that within the 
measurement uncertainty, the 21-cm detection rate in strong Mg~{\sc ii} systems is 
constant over $0.5<z<1.5$.  
Since stellar feedback processes are expected to diminish the filling factor of CNM 
over $0.5< z <1$, this lack of evolution in the 21-cm detection rate in Mg~{\sc ii} 
absorbers is intriguing.
Further, we find that if majority of 21-cm absorbers arise from DLAs then the 
cross-section of 21-cm absorbing gas i.e. cold neutral medium amongst DLAs has 
increased from $z=3.5$ to $z=0.5$.  
In a sample of 13 $z>2$ DLAs with both 21-cm and H$_2$ (another tracer of cold gas)  
absorption measurements, we report two new H$_2$ detections and 
find that in 8/13 cases neither 21-cm nor H$_2$ 
is detected. This confirms that the H~{\sc i} gas in $z>2$ DLAs is  
predominantly warm.  Interestingly, there are two cases where 21-cm absorption is not detected despite 
the presence of H$_2$ with evidence for the presence of cold gas.  This can be 
explained if H$_2$ components seen in DLA are compact ($\le$15\,pc) and contain 
$\le$10\% of the total N(H~{\sc i}).  
We briefly discuss results from our ongoing survey to identify 21-cm absorbers at low-$z$ 
to establish connection between 21-cm absorbers and galaxies, and constrain the extent 
of absorbing gas.  
\end{abstract}


\section{Introduction}\label{s:intro}

It is well known that physical conditions in the diffuse interstellar medium (ISM) of 
galaxies are influenced by various radiative and mechanical feedback processes associated 
with in-situ star formation. Therefore, volume-filling factors of different phases of gas 
in a galaxy are expected to depend on its star formation history. Of particular interest 
is the evolution of the volume-filling factor of cold neutral medium (CNM) phase that 
also serves as a gaseous reservoir for star formation in galaxies. 
Systematic searches of high-$z$ intervening 21-cm absorbers in samples of Mg~{\sc ii} 
systems and damped Ly$\alpha$ systems (DLAs) towards QSOs to measure CNM filling 
factor of galaxies have resulted in detections of 21-cm absorption towards $\sim$10-20\% 
of Mg~{\sc ii} systems and DLAs. However, establishing a connection between the redshift evolution 
of 21-cm absorbers and global star formation rate (SFR) density is not straight forward 
due to (i) small number statistics of 21-cm absorbers; (ii) ambiguities regarding the 
origin of absorbing gas; and (iii) issues related to the small scale structure in absorbing 
gas and the extent of radio source.
Here we present a summary of our recent results to constrain the evolution of cold gas 
in galaxies using 21-cm absorption and address issues related to partial coverage of 
absorbing gas.   

\section{21-cm absorption detection rate and CNM filling factor}

\begin{figure}
\centerline{
\hbox{
\includegraphics[width=5cm]{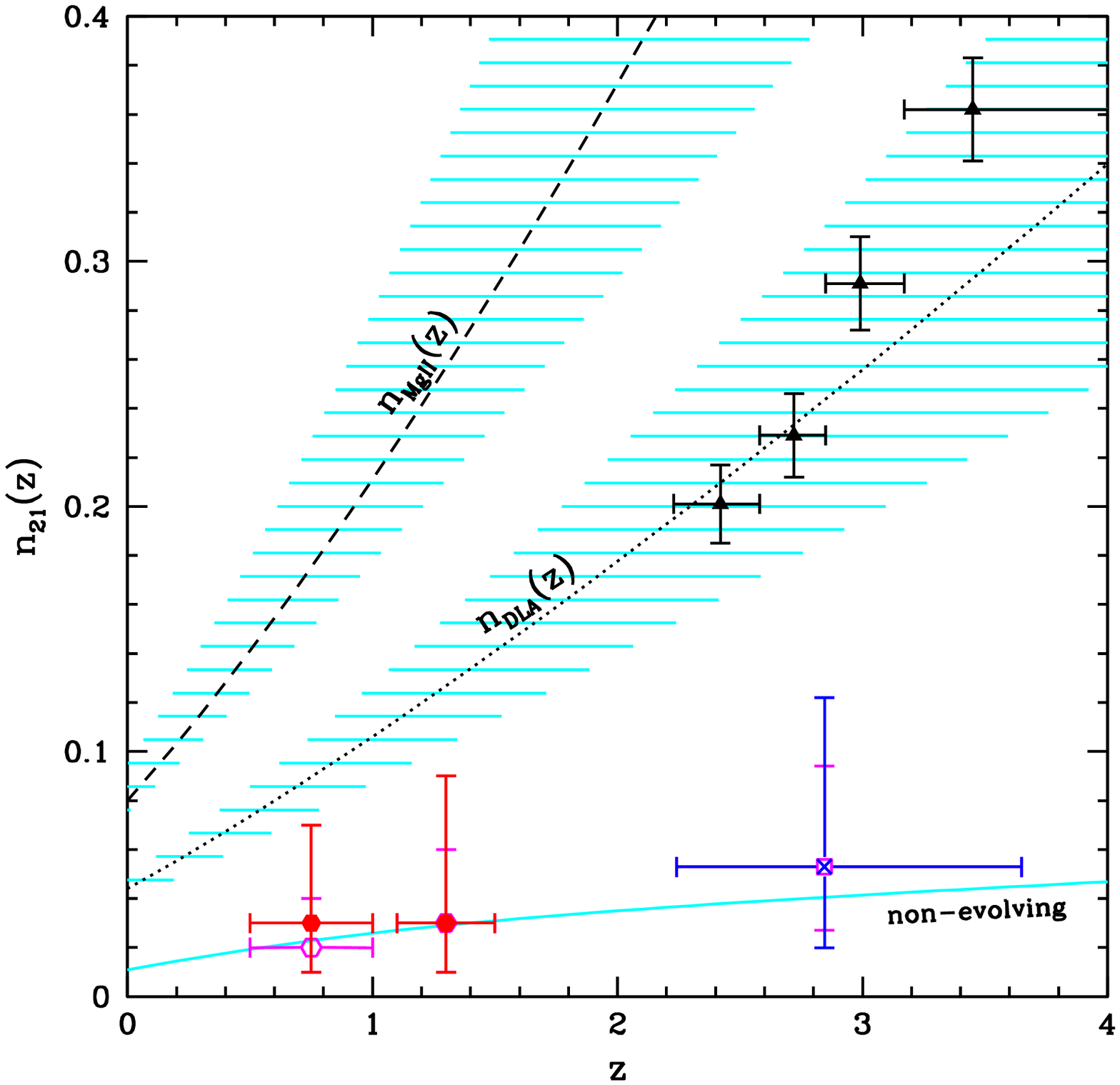}
\includegraphics[width=5cm]{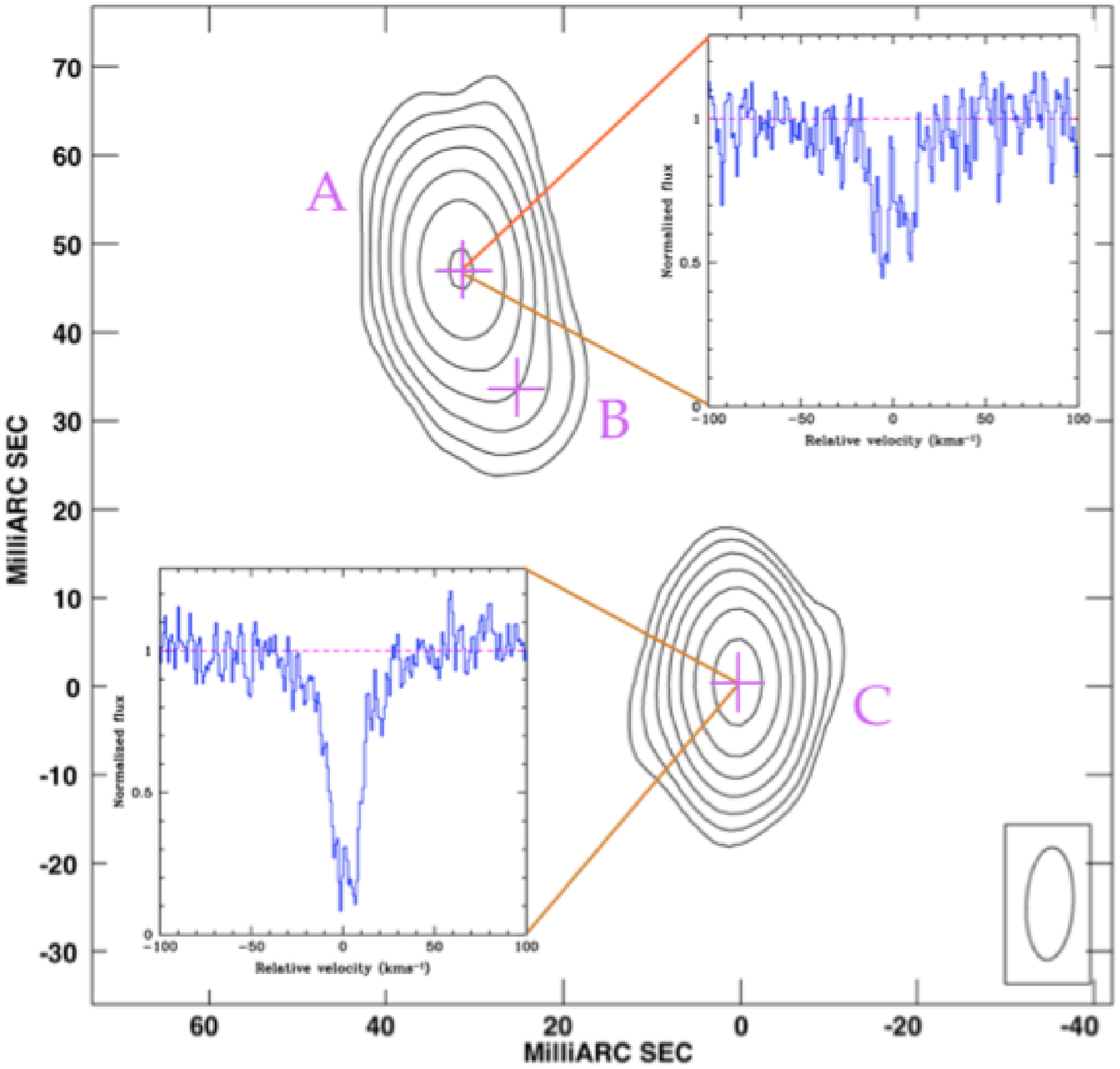}
}}
\caption{ {\bf Left:} Number of 21-cm absorbers per unit redshift range, n$_{21}$ ($z$), for integrated 21-cm 
optical depth sensitivity of 0.3\,km\,s$^{-1}$. The measurements at $z<2$ (circles) are based on strong 
Mg~{\sc ii} absorbers whereas at $z>2$ (square/cross) on DLAs. Open circle(s) and square are 
for no correction for partial coverage of background radio source by absorbing gas. 
Filled circle(s) and cross correspond to n$_{21}$ after the optical depths have been corrected 
for partial coverage using milliarcsecond scale radio maps (see Gupta et al. 2012 for details).  
The curve for non-evolving population of 21-cm absorbers normalized at n$_{21}$(z = 1.3) is also 
plotted. Lines and dashed areas show the number of absorbers per unit redshift for DLAs 
(dotted line) and strong Mg~{\sc ii} absorbers (dashed line). 
Triangles are the number per unit redshift of DLAs measured from the DLA sample based on SDSS Data Release 9. 
{\bf Right:} 
VLBA 21-cm absorption observations of $z=0.079$ absorber towards $z=0.993$ 
QSO J163956.35+112758.7 (Srianand et al. 2013). The background radio source is resolved 
into three components (A, B and C) with a maximum projected separation of 89\,pc. The 
integrated H~{\sc i} optical depth towards the northern and southern peaks 
(A and C respectively) in VLBA continuum map are higher than that measured in 
our GMRT spectrum and differ by a factor 2. 
} 
\end{figure}

The availability of large samples of low resolution quasar spectra from the Sloan Digital Sky 
Survey (SDSS) has led to compilations of large homogeneous catalogues of Mg~{\sc ii} absorbers 
at $z<2$ and DLAs at $z>2$.  We have been carrying out systematic searches of 21-cm absorption 
in samples of strong Mg~{\sc ii} systems (rest equivalent width$>$1\AA) and DLAs towards 
radio sources brighter than 50\,mJy at 
1.4\,GHz. These searches using the Green Bank Telescope (GBT), the Giant Metrewave Radio Telescope 
(GMRT) and the Westerbork Synthesis Radio Telescope (WSRT) have resulted in detection of 13 new 
21-cm absorbers at $0.5<z<2$ (Gupta et al. 2007, 2009, 2012) and one new absorber at $z=3.2$ 
(Srianand et al. 2010).  We have also complemented these 
observations with milliarcsecond scale imaging at 20-cm using Very Long Baseline Array (VLBA) 
to correct for the partial coverage of absorbing gas.  
Based on 21-cm optical depth measurements of $\sim$75 sight lines from our survey  
and $\sim$55 sight lines from the literature, we have constrained number per unit redshift 
range of 21-cm absorbers (n$_{21}$) at $z<3.5$ (Fig.~1). 
We find that within the measurement uncertainty, the 21-cm detection rate in strong Mg~{\sc ii} 
systems is constant over 0.5$<z<$1.5, i.e., over $\sim$30\% of the total age of universe. 
Since stellar feedback 
processes are expected to diminish the filling factor of CNM over 0.5$<z<$1, this lack of evolution 
in the 21-cm detection rate in strong Mg~{\sc ii} absorbers is intriguing.

Since the fraction of strong Mg~{\sc ii} absorbers that are likely to be DLAs is known from 
HST surveys at $z<2$, it is possible to determine 21-cm detection rate amongst DLAs at $z<2$.
We find that the 21-cm absorption detection rate in the DLAs at $z>2$ is lower by at most a 
factor 3 compared to the upper limits we obtained using the strong Mg~{\sc ii} absorbers at $0.5 < z < 1.0$. 
If the majority of 21-cm absorbers arise from DLAs then this would imply that the cross-section of 
the 21-cm absorbing gas i.e. CNM amongst DLAs has increased from $z = 3.5$ to $z = 0.5$ 
(Gupta et al. 2012).

\section{21-cm absorption and H$_2$}
As the interpretation of 21-cm absorption measurements is often complicated by more than 
one unknown amongst spin temperature, N(H~{\sc i}) and covering factor, it is useful to 
constrain physical conditions in absorbers with 21-cm absorption measurements using 
other tracers of cold gas such as H$_2$.
In our sample of 28 DLAs at $z>2$ presented in Srianand 
et al. (2012), we find that there are 13 DLAs for which  high-resolution optical 
spectra covering the expected wavelength range of H$_2$ absorption are available. 
We report the detection of H$_2$ molecules in the $z$= 3.4 21-cm absorber towards 
J0203+1134 (PKS 0201+113; Srianand et al. 2012), and $z=3.2$ absorber towards J1337+3152 
(Srianand et al. 2010). 
In eight cases, neither H$_2$ (with molecular fraction f(H$_2$)$\le$10$^{-6}$) nor 
21-cm absorption is detected. The lack of 21-cm and H$_2$ absorption in these systems 
can be explained if most of the H~{\sc i} in these DLAs originates from low-density 
high-temperature gas. In one case we have a DLA with 21-cm absorption not showing H$_2$ 
absorption.  In the remaining two cases 21-cm absorption is not detected despite 
the presence of H$_2$ with evidence for the presence of cold gas. All this combined with 
the constraints on density of gas from [C~{\sc ii}]* implies that 
the H$_2$ components seen in DLAs are compact (with sizes of $\le$15\,pc) 
and contain only a small fraction (i.e. typically $<$10 per cent) of the total 
N(H~{\sc i}) measured in the DLAs (Srianand et al. 2012).

\section{Partial coverage of background radio source}
In a sample of 54 quasars with 21-cm absorption measurements at $z < 2$ and milliarcsecond-scale VLBA 
images, we find 70\% of 21-cm detections to be towards quasars with linear sizes $<$100\,pc 
(Gupta et al. 2012). 
This suggests that 21-cm absorbing gas has a typical correlation length of $\sim$100\,pc.  
We use VLBA images to correct 21-cm optical depths for partial coverage. 
We find that the 21-cm detection rates can be underestimated by up to a factor 2 
if 21-cm optical depths are not corrected for the partial coverage. 

Unfortunately due to the unavailability of suitable low-frequency ($<$1\,GHz) receivers 
at VLBI stations it is not possible to perform milliarcsecond-scale spectroscopy and 
directly constrain the extent of absorbing gas. 
To summarize, milliarcsecond-scale 21-cm absorption spectroscopy has been done only in 5 cases. A 
lower limit of $\sim$2-30\,pc has been inferred for the absorber sizes from these observations. 
For example, see the case of $z=0.079$ absorber towards J1639+1127 shown in Fig.\,2 
(Srianand et al. 2013).  
A systematic large survey to identify new 21-cm absorbers at low-redshift to 
establish the connection of 21-cm absorbers with galaxies (Gupta et al. 2010, 2013) 
and suitable for milliarcsecond-scale spectroscopy with VLBA is in progress.  
Finally, 21-cm detections of the order of DLA sample based on SDSS Data Release 9 will 
be possible with our MeerKAT Absorption Line Survey (MALS). Blind searches of 21-cm absorption 
lines with MeerKAT and other SKA pathfinders will provide a complete view of the evolution of cold gas 
in galaxies and shed light on the nature of Mg~{\sc ii} systems and DLAs, and their relationship 
with stellar feedback processes.

\section*{Acknowledgements}

We thank GBT, GMRT, VLBA and WSRT staff for their support during the observations 
of data used in the work presented here. 
We also acknowledge the use of SDSS spectra and images.  


\end{document}